\documentclass[runningheads]{svmult}

\usepackage{makeidx}   
\usepackage{graphicx}  
\usepackage{subeqnar}  
\usepackage{multicol}  
\usepackage{physprbb}  
\makeindex             

\newcommand{\msolgg} {M$_{\odot}$}

\begin{document}
\title*{Early Virtual Science: some lessons for the AVO}

\toctitle{Early Virtual Science: some lessons for the AVO}
\titlerunning{Early Virtual Science: some lessons for the AVO}
%
\author{Gerry Gilmore}

\authorrunning{Gerry Gilmore}

\institute{Institute of Astronomy, Madingley Rd, Cambridge CB3 0HA, UK}

\maketitle              

\begin{abstract}

Experience with ASTROVIRTEL, scientific analysis of current large data
sets, and detailed preparation for the truly huge future missions
especially GAIA, provide important lessons for the Astrophysical
Virtual Observatory. They demonstrate that the science cases are
impressive, specifically allowing new thresholds to be crossed. The
AVO is more than just faster cheaper better, it allows the new.  The
example of use of pre-explosion imaging of supernova to identify
progenitors is used to illustrate some general challenges.  Some
non-trivial technical astronomical issues arise, especially
astrometry, to complement the many technical implementation
challenges.  A critical scientific lesson is the need to quantify data
quality. How are we to ensure the Virtual Observatory produces top
science, and avoids being overwhelmed with mediocre data?

\end{abstract}

\section{Introduction}

The Astrophysical Virtual Observatory has an exceptionally strong
science case. One example, identification of the pre-explosion state
of core-collapse supernovae, is described below. This case illustrates
some of the technical and strategic challenges which the Virtual
Observatory projects have still to face.

There is a strong scientific
case to identify and access appropriate archival data on the sites of
supernova explosions. Since the candidate star is identified by its
self-immolation, the only relevant data are archival!  A project to
obtain suitable new data to act as a future archive is underway with
HST. In addition, much valuable data already exists. Some of this is
in archives, and is well calibrated. Much is in private hands. The
effort needed to use even excellent quality calibrated and published
data is illustrated below, using as examples the searches by Smartt
etal for the progenitors of SN1999em and SN2002ap.

One the most difficult of the astronomical challenges facing the
integrated use of federated multi-wavelength multi-resolution archives
involves source extraction, and astrometric cross-matching. Some examples
are given below. The challenge is however obvious to all who have
experience even with combination of HST WFPC2 and NICMOS images. Even
in this case, with high-quality, high spatial resolution, stable,
well-calibrated data sets, with only a one-half decade wavelength
range, simple matching of optical and near-IR images of star clusters
is not trivial. A nice example is available in figures one and two of 
Johnson etal (2001), who present PC and NICMOS images of two young LMC
globular clusters.

A second challenge involves data reliability. The example below
illustrates how even unusually high-quality data cannot be used beyond
the range in which their systematic uncertainties become relevant to
the science at hand. However, few (if any) data archives are
calibrated well enough to provide this information, except in response
to very specific questions and applications.

This raises the spectre of well-meaning providers of what are in fact
data of limited calibration ensuring that virtual observatory data
product users either produce defective science, or are overwhelmed
with learning the limits of every individual data set accessed by the
entire system. Might it be that data-archives need a
quality-assessment check before they are `eligible' for access? Or is
it to be {\em caveat emptor?}

\section{An Example Application: Identifying the progenitors of
Type~II supernovae}

Supernovae are the evolutionary end points of all stars more massive
than about 8\msolgg. Predictions of the pre-explosion evolutionary
status of these stars is a key test of stellar evolutionary theory.
Supernova explosions additionally drive the chemical evolution of the
Universe and play a major role in shaping the dynamics of the
interstellar medium of gas rich galaxies. They are of crucial
importance to fundamental studies of the evolution of galaxies and
the origins of the chemical elements in the Universe.

The spectra of supernovae come in many different varieties, with the
classifications based on the lines observed and the temporal evolution
of these features. The presence of broad H\,{\sc i} optical lines
indicates a SN Type\,II classification, while those that do not show
hydrogen are classed Type\,I. The SNe\,Ia are thought to arise through
thermonuclear explosions in white dwarf binary systems, hence the
progenitors are low-intermediate mass stars. All other supernovae
including the Types Ib/Ic and all flavours of Type II are thought to
be due to core-collapse during the deaths of massive stars. SNe\,II
show prominent, broad H\,{\sc i} lines in their optical spectra,
indicating that the progenitor retained a substantial hydrogen
envelope prior to explosion. SNe\,Ib/Ic do not show any significant
signs of hydrogen in their spectra, although SNe Ib display pronounced
He\,{\sc i} absorption. 

There is  strong though indirect evidence that  SNe~II
and SNe~Ib/Ic are associated with the deaths of massive stars, as they are
never seen in elliptical galaxies, are observed only rarely in S0
galaxies, and they often appear to be associated with sites of recent
massive-star formation, such as H\,{\sc ii} regions and OB associations
in spiral and irregular galaxies \cite{vandyk96,fili97}.  The Type\,II
events are further seen to be split into subtypes (IIb, IIn, II-L and
II-P). Leonard etal \cite{leo2001} discuss the widely held belief that
core-collapse events can be ranked in order of their increasing
hydrogen envelope mass at the time of explosion, which is - Ic, Ib,
IIb, IIn, II-L, II-P.

This overwhelming, but still indirect, evidence implies that SNe\,II
arise from the deaths of single, massive stars, with initial masses
M$> 8-10$M$_{\odot}$, which have retained a substantial fraction of
their hydrogen envelope.  However there has been only one definite and
unambiguous detection of a star that has subsequently exploded as a
supernova of any type $-$ that of Sk$-69^{\circ}202$, the progenitor to
SN1987A in the LMC \cite{white87}. Prior to explosion this star was a
blue supergiant of B3\,Ia spectral type \cite{wal89}, which would
correspond to $T_{\rm eff} =18000$\,K (from the temperatures in
\cite{mcer99}) and $\log L/L_{\odot} = 5.1$ (from the photometry in
\cite{wal89}), and an initial mass of $\sim$20M$_{\odot}$. The closest
supernova to the Milky Way since then was SN1993J in M81 (3.63\,Mpc),
which was a Type\,IIb event.  Ground based $UBVRI$ photometry of the
SN site before explosion was presented by Aldering etal\cite{alder94}. The
photometry of the progenitor candidate was best fit with a composite
spectral energy distribution of a K0\,Ia star and some excess UB-band
flux suspected to be from unresolved OB association contamination
(confirmed by recent HST observations).  Neither
the progenitor of SN1987A nor that of SN1993J is consistent with the
canonical stellar evolution picture, where core carbon burning
finishes and core-collapse occurs relatively soon afterwards
($\sim10^3-10^4$\,yrs) while the massive star is an M-supergiant.

Other attempts have been made to identify SNae
progenitors on pre-explosion archive images, with little success in
directly detecting progenitor stars.  Fortunately, even an upper limit
on progenitor brightness provides a mass-limit. An upper mass limit to the
progenitor of SN1980K has been estimated to be $\sim$18M$_{\odot}$
\cite{thom82}, while only an upper limit to the absolute visual
magnitude was determined for SN1994I \cite{barth96}.  Recently
we \cite{smartt2001} studied HST archive images of the site of the
Type\,II-P SN1999gi which were taken before explosion.  SN1999gi
occurred in a young OB-association: however the the progenitor was
below the detection limit of the available pre-explosion images. By
determining the sensitivity of these exposures and comparing the
estimated bolometric luminosity with stellar evolutionary theory, an
upper limit to the mass of the progenitor was set at
9$^{+3}_{-2}$\msolgg.

\section{SN1999em: archival experience with excellent published data}

SN1999em in NGC1637 was discovered on Oct. 29 1999 by the Lick Observatory
Supernova Search  \cite{li99} at an unfiltered CCD magnitude
of $\sim13.5^m$. It was soon confirmed to be a Type\,II and being a
very bright event it has been studied extensively in the optical since
then. It has been firmly established as a normal Type II-P event,
having a plateau phase lasting approximately 90 days after discovery
\cite{leo2001}.  There have also been UV, X-ray, radio, and
spectropolarimetry observations. \cite{baron2000} have presented model
atmosphere fits to the early-time optical and HST-UV spectra,
indicating that an enhanced He abundance is required to fit the data
satisfactorily. They further use the very blue continuum of the early
spectrum to determine a reddening.  The expanding photosphere method
(EPM) has been applied to SN1999em by Hamuy etal\cite{hamuy2001} to determine a
distance to the host galaxy of $7.5\pm0.5$\,Mpc, illustrating the
possibility of using SNe\,II-P as luminous distance
indicators. Chandra and radio observations of SN1999em have been used
to probe the interaction of the SN ejecta with the circumstellar
material, which are consistent with a mass-loss rate of
$\sim2\times10^{-6}$\msolgg\,yr$^{-1}$ and a slow  wind
\cite{pool2001}.  Given the substantial interest in this bright
supernova and the extensive multi-wavelength observations of the event
it is of great interest to have direct information on the progenitor
star.  Further it would be desirable to have more detections of
progenitor stars (as in SN1987A) in order to draw a meaningful
physical picture of what causes the different varieties of
core-collapse events.

\begin{figure}[ht]
\begin{center}
\includegraphics[width=\textwidth]{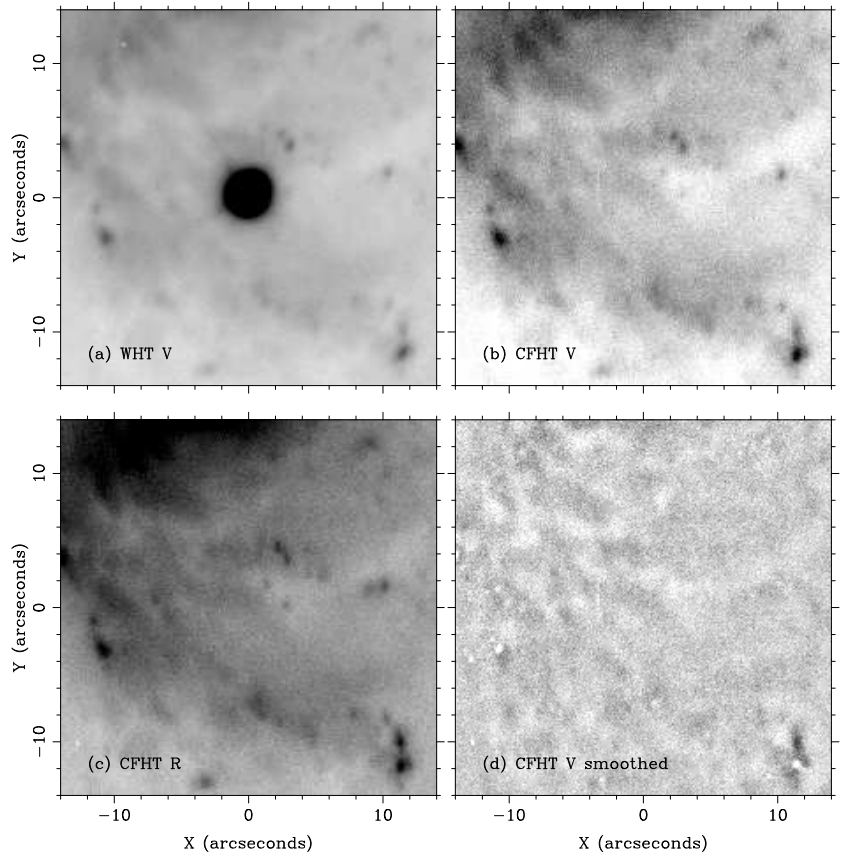}
\end{center}
\caption{
{\bf (a):} The position of SN1999em in NGC1637 in a transformed WHT
post-explosion image. In this image the centroid of the SN is
saturated but a further short exposure is used to measure it
accurately, and is set at (0,0) in all frames. {\bf (b)} and {\bf
(c):} The region of pre-explosion $VR$-band CFHT images. {\bf
(d):} An image with a smooth background removed and all PSFs from
single stars subtracted.  Sohn \& Davidge (1998) catalogue a star with
a coordinate of $(0.08'',-0.24'')$, and magnitude $V=23.47, R=23.33$
which is within the astrometric error of the transformation discussed
in the text. However on close inspection there is no evidence
for a point source at this position in any of the $VRI$ bands. The two
stars at (1.6, 1.9) and (2.9,0.2) are have $V=23.97, 23.15$ and
$V-R=0.26, 0.10$ respectively. The detection limit is position
dependent as the background varies considerably over small scales.
This figure is from Smartt, Gilmore, Tout \& Hodgkin 2002a.}
\label{sn_closeup1}
\end{figure}

By chance there are optical images of this galaxy taken 7 years before
SN1999em occurred in the archive of the Canada France Hawaii
Telescope, maintained at the Canadian Astronomy Data
Centre\footnote{http//:cadcwww.dao.nrc.ca/cfht/}.  These
high-resolution images were taken by Sohn \& Davidge\cite{sohn98},
who presented photometry of the luminous supergiant members of the
galaxy. Amongst other results in their paper, a distance of 
7.8$\pm1$\,kpc is derived from the magnitudes of the galaxy's brightest
stars. As SNe Type\,II are thought to have luminous supergiant
progenitors, high quality pre-explosion images of nearby
galaxies which resolve the brightest stars could allow direct
detection of progenitors, or at least limits to be set on luminosity
in the event of a non-detection. 

In fact, Sohn \& Davidge published
photometry of a source coincident with the supernova position.
Smartt etal (2002a) however presented an accurate
astrometric determination of the position of SN1999em on the
pre-explosion frames. They show that there is no detection of a point
source at this position which could be interpreted as the progenitor.
The detection limits of the exposures are determined, allowing
bolometric luminosity limits and an upper mass limit to be determined
for the progenitor star. 

This discrepancy between different analyses of the same high-quality
data using essentially the same photometric techniques illustrates one
of the primary challenges for the Virtual Observatory: even
high-quality data are often not suitably calibrated for use in a
different application than that for which they were obtained, without
considerable interactive analysis by an experienced astronomer.

\subsection{The challenges of repeat Data Analysis}

The galaxy NGC\,1637 was observed on 5th January 1992 on the CFHT with
the HRCam \cite{mc89}, with exposures of 900s, 750s
and 600s in $V, R_{\rm C}, I_{\rm C}$. The material is publicly
available through the CFHT archive at CADC$^{1}$.  The reduction,
analysis and multi-colour photometry of the bright stellar objects in
the field was presented by \cite[hereafter SD98]{sohn98}.
 Their limiting magnitudes for detection, defined as the
magnitude where DAOPHOT \cite{stet87} predicts errors of
$\pm0.5$ or greater, are $\sim$24.9, 24.8, 23.9 in $V, R_{\rm C},
I_{\rm C}$ respectively. These data hence probe stars brighter than
M$_{v}\simeq-4.9$, assuming the distance modulus from SD98, and their
estimates for average line of sight extinction of $A_V = 0.34$. The
image quality of the archive data are $0.7''$ FWHM in all three
bands. SD98 determined the colours for 435 objects in the frames which
are simultaneously detected in all three filters.  The CFHT HRCam used
a 1024$\times$1024 pixel Ford-Aerospace CCD mounted at prime focus,
with 18$\mu$m pixels, corresponding to $0.13''$ on the sky.

On 28th November 1999, Smartt etal obtained two $V-$band images of
NGC1637 on the William Herschel Telescope on La Palma $-$ 30 days
after discovery of SN1999em. The AUX-port camera at Cassegrain was
used, which has a 1024$^{2}$ Tektronix detector (ING CCD TEK2) at a
plate scale of $0.11''$\,pix$^{-1}$. This was done through the ING
{\sc service} program and two exposures were taken (900s and 10s),
during which the seeing was $0.7''$ FWHM.
The considerable similarities between the cameras, telescope apertures and
observing conditions in both cases mean the sensitivities of the pre and
post-explosion data are very similar, and substantially ease joint
analysis (see Fig~1).

Astrometrically calibrating either of the two frames as they stand
onto an absolute reference frame is not possible due to their limited
FOV, and the fact that any isolated stars outside the main body of the
galaxy which could be used as secondary astrometric standards are
saturated in the deep CCD frames. However, given the similarity in the
plate scales and the detection limits of the two data sets, Smartt
etal were able to perform a simple geometric transformation of the WHT
pixel array onto the CFHT array (similar to the method in Smartt et
al. 2001a). First of all they identified ten bright, relatively
isolated stars in both the WHT 900s $V$ exposure and CFHT $V$ frame,
and measured the centroids of the stars on the WHT frame by fitting a
model point-spread-function (PSF) to each using standard techniques in
{\sc DAOPHOT} within {\sc IRAF}, using the pixel coordinates of the 10
stars from the tabulated photometry of SD98. A spatial transformation
function was calculated, which fitted a linear shift, a magnification
factor and a rotation angle.  Polynomials of various orders were tried
to fit the $x$ and $y$ mapping, but the results were no better than
the simple scaling formerly described. The transformation function was
applied to the WHT 900s frame, and both were trimmed to the common
region of overlap (625$\times$560 pixels, as shown in Fig~1. This
process left no residual systematic difference in the pixel astrometry
between these two datasets.  The mean offset in radial positions
of the stars in the CFHT and WHT frames is $\delta r = 0.17''
\pm0.13$.

This astrometric mapping process, already rather more elaborate than
could be achieved by even careful use of WCS FITS headers, was further
complicated by the wide dynamic range difference between the bright
and the faint stars, which exceeds the detector dynamic range.  The
supernova itself of course was saturated, and so would be rejected
from analysis by most pipeline processing systems. Thus a further
short-exposure astrometric transfer image was also required.  Using
the stellar centroid method to check for offsets between the long and
short frames proved problematic due to the low counts in stars in the
short exposure frame; a significantly longer exposure would have led
to saturation of the SN. However some stars in common could be
matched, and indicated mean offsets of ($-0.01''$,$0.01''$).

All these transformations required care and one-off interactive checking.

\subsection{Re-use of published photometry}

The photometry list of SD98 reports the detection of star \#66
(hereafter NGC1637-SD66) at $(0.08'',-0.24'')$ and the nearest other
object is $2.5''$ away. Star NGC1637-SD66 is the only candidate for
the progenitor in the existing photometry of SD98, at a distance of
$\delta r = 0.25''$ from SN1999em.  This does fall within the
$1\sigma$ standard deviation of the differences in positions of the
106 matched stars, and hence is compatible with being coincident with
the supernova position.

However on closer inspection this does not appear to be a reliable
detection of a stellar-like object. In Fig~1 the region around
SN1999em is displayed from the CFHT $V$ and $R$ band images. There is no
obvious resolved luminous object from a visual inspection and star
NGC1637-SD66 is not apparently obvious (the results for the $I$-band
data are similar).  The position of the supernova appears to lie on a
faint ``ridge'' (running diagonally left-right in the figure), and the
detection limits of the image are highly position dependent given the
variable background. In deriving their final photometric list, SD98
applied a background smoothing technique to recover faint stars
against the varying galaxy background.  Smartt etal repeated this method
to determine if any sign of a single point source at the SN1999em
position appears after background subtraction, following the steps
described in \cite{sohn96}. The {\sc daophot} package was used to fit
model PSFs to the brighter stars in the images. These were subtracted
from the data and a boxcar median filter of pixel dimension
25$\times$25 (i.e. 5 times the seeing width) was applied to this
subtracted image. This was assumed to be indicative of the varying
background of the galaxy and was subtracted from the original
frame. The PSF fitting routines within {\sc daophot} were re-run on
the resultant frame. The results from this for the $V$ band are shown
in Fig~1(d), where the point sources subtract off quite cleanly apart
from some objects which are not resolved but are broader than a PSF.
Again there is no clearly identifiable point-source at the SN1999em
position after the smoothing technique is applied, and no object is
visible in the $R$ and $I$ frames either.

Smartt etal conclude that the progenitor of SN1999em is below the
sensitivity limits of the pre-explosion $VRI$ data, and that the star
NGC1637-SD66 detected by Sohn \& Davidge is a noise fluctuation which
survived even their exceptionally careful photometric analysis.

\section{SN2002ap: a more complex set of archive data}

Supernova~2002ap was discovered by Yoji Hirose on 2002 January 29.4 UT
in the spiral galaxy M74 \cite{nakano02}.  It was discovered at
$V=14.54$, and at a distance of approximately 7.3\,Mpc, may be the
closest supernova since SN~1993J in M81 (at 3.6\,Mpc). Several
observers rapidly obtained spectra, reported that it appeared similar
to the peculiar SN~1998bw but caught at an earlier epoch
\cite{meikle02}, exciting much activity. Later optical spectra of
SN~2002ap indicate that it does appear to be a Type Ic, and its
optical lightcurve appears to have peaked at approximately $M_V \simeq
-17.5$, some 1.7$^{m}$ fainter than SN~1998bw. Two popular theories
for the origin of Type\,Ic supernovae are the core collapse of massive
stars when they are in the WR phase, or the core collapse of a massive
star in an interacting binary which has had its envelope stripped
through mass transfer.


\vskip 5truecm
{GIF format figure 2

SN2002ap in M74:  prediscovery optical images, with
$BVR{\rm H}\alpha$ from the KPNO\,0.9m and $UI$ from the INT WFC. The
location of SN~2002ap is at the centre of each frame, indicated by the
orthogonal lines. The SN position is $2.31'' \pm 0.29''$ away from the
nearby bright object detected in $BVRI$ (and marginally seen in $U$)
i.e. it is clearly not coincident with this source. This figure is
from Smartt etal 2002b.}

The host galaxy, M74, is a large and pretty spiral, which has been
much imaged by many telescopes, including HST and
Gemini-North. However, the supernova lies {\em JUST OFF} the field of
view of all these high-quality studies. Archival analysis did identify
one set of wide-field optical images of M74 taken before discovery of
2002ap. These images are from the Wide-Field-Camera (WFC) on the Isaac
Newton Telescope (INT), La Palma, taken on 2001 July 24 through
filters $UBVI$. The exposures were 120s in each of $BVI$ and 180s in
$U$. These were taken at the end of a night during the Wide Field
Survey programme on Faint Sky Variability \cite{groot02}. The WFC
comprises 4 thinned EEV 4k$\times$2k CCDs, with 13.5$\mu$m ($0.33''$)
pixels. Repeat exposures of 120s in $UVI$ were taken on 2002 February
2. The supernova core saturated in these frames, and shorter 2-10s
exposures were taken with the telescope guiding continuously between
the short and long exposures to determine an accurate position for
SN~2002ap.

A second set of images was identified by searching the ADS for
published studies of M74 which would indicate the existence of data,
even though those data had not been archived. Such a study was
identified , using data from the KPNO 0.9m with the Direct Imaging
Camera taken on 1993 September 15 \& 17.  The authors were contacted,
the tapes found, and {\em ad hoc} virtual access turned into real
access.  The subsequent analysis of these data followed the
methodology outlined above, and is detailed by Smartt etal 2002b.
Figure~2 shows the outcome, where yet again no progenitor was
detected...

The galaxy M74 has been imaged by HST, Gemini, CFHT and WHT, however
the supernova position does not fall on any of these images.  All the
publicly available archives have been searched for deeper, higher
resolution images of M74 but there are no superior images to those
shown in figure~2 that include the pre-explosion site of SN~2002ap.

The important lesson for the present is that truly `virtual' data, ie
data whose existence can be deduced but which are not even physically
in an archive,  can prove of considerable science value.

\section{ Supernova progenitors: the Future Virtual Observatory- real
observatory interface}

We require data on more progenitors before we can be confident of the
origins of the core-collapse SNe sub-types. Prompt and frequent
multi-wavelength observations of SNe provide quite detailed
information on the explosion and circumstellar material, and by
inference on the mass-loss and envelope properties of the
progenitor. However having high-quality archive images of SNe sites
taken {\em prior to explosion} is the only robust way to set firm
limits on the nature of the progenitor stars.

\begin{table}[h]
\caption{Comparison of all information that is currently available
from direct observations of the progenitors of core-collapse SN.
The metallicity refers to estimates for the progenitor star, in 
the case of the spiral galaxies from measured abundance gradients
and galactocentric radii of the SN. Mass refers to the {\em main-sequence} 
mass of the progenitor.\label{sncomp}}
\begin{center}
\begin{tabular}{lllll}
\hline\noalign{\smallskip}
SN      & Type           & Mass       & Z & Spec. Type \\ 
\hline
1987A   & II peculiar    & ~~~20\msolgg   & 0.5Z$_{\odot}$ &  B3Ia  \\
1980K   & II-L           & $<20$\msolgg & 0.5Z$_{\odot}$     &  ? \\
1993J   & IIb            & ~~~17\msolgg & $\sim$2Z$_{\odot}$  &  K0Ia \\
1999em  & II-P           & $<12$\msolgg & 1$-$2Z$_{\odot}$ & M-supergiant ? \\
1999gi  & II-P           & $<9$\msolgg  & $\sim$2Z$_{\odot}$ & M-supergiant ? \\
2002ap  &  I-c           & ~~~25\msolgg &  0.5Z$_{\odot}$ &  WR? binary?\\
\noalign{\smallskip}
\hline
\end{tabular}
\end{center}
\end{table}

Observations of nearby spiral and irregular galaxies within
$\sim$20\,Mpc of the Milky Way allow the massive stellar content to be
resolved. Multi-band images from the Hubble Space Telescope of all the
face on spirals would be an excellent archive for future use when SNe
are discovered. In the worst case this will allow limits to be set on
the progenitor masses, as shown here and in Smartt et al. (2001), and
should lead, in some cases, to definite identifications of progenitor
stars. Already the HST archive contains approximately 120 Sb-Sd
galaxies within $\sim$20\,Mpc which have observations of useful depth
in at least 2 broad-band filters.  There are a further 130 Sb-Sd
spirals with exposures in 1 broad-band filter.  Smartt etal have a
Cycle\,10 HST project to supplement the latter 130 galaxies with 2
further filters, and observe 120 more late-type spirals in three
filters.  This should give a total of $\sim$370 Sb-Sd galaxies with
HST observations. This number is steadily increasing each year, with
data coming from projects with other scientific goals.  This is
supplemented with high-quality ground-based images from the well
maintained archives of the ESO, ING, CFHT (and soon Gemini).

The various initiatives aimed at producing combined virtual
observatories have, amongst many other applications,  the unique
historical aspect which is essential to SNe progenitor searches. One
of the first of these ({\sc
astrovirtel}\footnote{http://www.stecf.org/astrovirtel}) has already
allowed us to search multi-telescope archives (HST + ESO telescopes)
and use catalogue data as search criteria (e.g. LEDA). Along with some
manual searching of the ING and CFHT archives, this suggests there are
a further 100 spirals with ground-based observations of the quality
presented in figures 1 and 2. Assuming a combined SNe II/Ib/Ic rate of
$1.00\pm0.4$\,$(100{\rm yr})^{-1}(10^{10}L^{B}_{\odot})^{-1}$
\cite{capp99}, and that the galaxies in our archive
have a mean luminosity $\sim10^{10}L^{B}_{\odot}$, then one would
expect $\sim4.7\pm2$ core-collapse SNe per year in this sample. As the
field-of-view of the WFPC2 on HST will only cover an average of 50\%
of the area of the optical disk of spirals between 10-20\,Mpc, then an
estimate of the number of SNe which will have pre-explosion archive
material available is $\sim2.4\pm2$ per year. Within a period of
$3-5$\,yrs we would hence expect the statistics presented in Table\,1
to improve significantly.  This is an example of unique science to be
done with future Virtual Observatories.

\section{Conclusions}

Identification and analysis of high-resolution ground-based images of the
pre-explosion sites of Type\,II supernovae provide unique
information on the late stages of evolution of massive stars, the
chemical evolution of the Universe, and the physics of feedback on
galaxy evolution. Examples of the successful application of this
method are presented above. From a Virtual Observatory perspective
there are some important lessons to be learned:

\begin{itemize}
\item Even excellent quality and carefully derived data products
(crowded field photometry in the specific example here) can prove
unsuitable for purposes different than their original application. The
scientific integrity of data retrieved by the Virtual Observatory
`system' must still be established by the responsible astronomer.
\item A considerable amount of non-archived non-calibrated data exist
in private repositories. While in some cases this can be of unique
value, the work involved in retrieving, and especially in calibrating,
old data suggests this is a
worthwhile use of resources only in special cases.
\item It is interesting to consider if the point above means that the
AVO should restrict itself to accessing only well-described and major
public data sets.
\item A general challenge is to provide adequate astrometric
cross-matching for different datasets. This raises the issue of
different spatial resolutions, source dropouts, etc. The examples
given here suggest that no complete general solution is feasible,
except for imaging data from telescopes with extremely well-quantified
optical systems.
\end{itemize}

\end{document}